\documentclass{article}
\usepackage{amsmath}
\usepackage{amsmath,amssymb,epsfig}

\begin{document}

\thispagestyle{empty} \vspace*{1cm} \rightline{Napoli DSF-T-31/2005} %
\rightline{INFN-NA-31/2005} \vspace*{2cm}

\begin{center}
{\LARGE Topological order and magnetic flux fractionalization in Josephson
junction ladders with Mobius boundary conditions: a twisted CFT description}

{\LARGE \ }

\vspace{8mm}

{\large Gerardo Cristofano\footnote{{\large {\footnotesize Dipartimento di
Scienze Fisiche,}\textit{\ {\footnotesize Universit\'{a} di Napoli
``Federico II''\ \newline
and INFN, Sezione di Napoli}-}{\small Via Cintia - Compl.\ universitario M.
Sant'Angelo - 80126 Napoli, Italy}}}, Vincenzo Marotta\footnotemark[1]  , }

{\large Adele Naddeo\footnote{{\large {\footnotesize Dipartimento di Scienze
Fisiche,}\textit{\ {\footnotesize Universit\'{a} di Napoli ``Federico II'' 
\newline
and INFM, Unit\`{a} di Napoli}-}{\small Via Cintia - Compl. universitario M.
Sant'Angelo - 80126 Napoli, Italy}}}, Giuliano Niccoli\footnote{{\large 
\textit{\footnotesize Sissa and INFN, Sezione di Trieste - Via Beirut 1 -
34100 Trieste, Italy}}} }

{\small \ }

\textbf{Abstract\\[0pt]
}
\end{center}

\begin{quotation}
We propose a CFT description for a closed one-dimensional fully frustrated
ladder of quantum Josephson junctions with Mobius boundary conditions \cite
{noi}; we show how such a system can develop topological order thanks to
flux fractionalization. Such a property is crucial for its implementation as
a \textquotedblleft protected\textquotedblright\ solid state qubit.

\vspace*{0.5cm}



{\footnotesize Work supported in part by the European Communities Human
Potential}

{\footnotesize Program under contract HPRN-CT-2000-00131 Quantum
Spacetime\newpage } \setcounter{page}{2}
\end{quotation}

\section{Introduction}

Arrays of weakly coupled Josephson junctions provide an experimental
realization of the two dimensional ($2D$) XY model physics. A Josephson
junction ladder (JJL) is the simplest quasi-one dimensional version of an
array in a magnetic field \cite{ladder}; recently such a system has been the
subject of many investigations because of its possibility to display
different transitions as a function of the magnetic field, temperature,
disorder, quantum fluctuations and dissipation. In this contribution we
address the phenomenon of fractionalization of the flux quantum $\frac{hc}{2e%
}$ in a fully frustrated JJL, the basic question being how the phenomenon of
Cooper pair condensation can cope with properties of charge (flux)
fractionalization, typical of a low dimensional system with a discrete ($%
Z_{2} $ in our case) symmetry. Then we discuss how it is deeply related to the issue of
topological order.

We must recall that charge fractionalization has been successfully
hypothesized by R. Laughlin to describe the ground state of a strongly
correlated $2D$ electron system, a quantum Hall fluid, at fractional
fillings $\nu =\frac{1}{2p+1}$, $p=1,2,...$. In such a system charged
excitations are present with fractional charge (anyons) and elementary flux $%
\frac{hc}{e}$. Furthermore the phenomenon of fractionalization of the
elementary flux has been found more recently in the description of a quantum Hall fluid at
non standard fillings $\nu =\frac{m}{mp+2}$ \cite{cgm2}\cite{cgm4}, within
the context of $2D$ Conformal Field Theories (CFT) with a $Z_{m}$ twist.

In Refs. \cite{noi1} it has been shown that the presence of a $Z_{2}$
symmetry accounts for more general boundary conditions for the propagating
electron fields which arise in quantum Hall systems in the presence of
impurities or defects. Furthermore such a symmetry is present also in the
fully frustrated XY (FFXY) model or equivalently, see Ref. \cite{foda}\cite
{noi}, in two dimensional Josephson junction arrays (JJA) with half flux
quantum $\frac{1}{2}\frac{hc}{2e}$ threading each square cell and accounts
for the degeneracy of the ground state.

It is interesting to notice that it is possible to generate non trivial
topologies, i.e. the torus, in the context of a CFT approach. That allows in
our case to show how non trivial global properties of the ground state wave
function emerge and how closely they are related to the presence of half
flux quanta, which can be viewed also as ``topological defects''.

The concept of topological order was first introduced to describe the ground
state of a quantum Hall fluid \cite{wen}. Although todays interest in
topological order mainly derives from the quest for exotic non-Fermi liquid
states relevant for high $T_{c}$ superconductors \cite{fisher}, such a
concept is of much more general interest \cite{wen1}.

Two features of topological order are very striking: fractionally charged
quasiparticles and a ground state degeneracy depending on the topology of
the underlying manifold, which is lifted by quasiparticles tunnelling
processes. For Laughlin fractional quantum Hall (FQH) states both these
properties are well understood \cite{wen2}, but for superconducting devices
the situation is less clear.

Josephson junctions networks appear to be good candidates for exhibiting
topological order, as recently evidenced in Refs. \cite{ioffe}\cite{pasquale}
by means of Chern-Simons gauge field theory. Such a property may allow for
their use as ``protected'' qubits for quantum computation.

The aim of this contribution is to show that the twisted model (TM) well
adapts to describe the phenomenology of fully frustrated JJL with a
topological defect and to analyze the implications of ``closed''\ geometries
on the ground state global properties. In particular we shall show that
fully frustrated Josephson junction ladders (JJL) with non trivial geometry
may support topological order, making use of conformal field theory
techniques \cite{noi}. A simple experimental test of our predictions will be
also proposed.

The paper is organized as follows.

In Section 2 we introduce the fully frustrated quantum Josephson junctions
ladder (JJL) focusing on non trivial boundary conditions.

In Section 3 we recall some aspects of the $m$-reduction procedure \cite{VM}%
, in particular we show how the $m=2$, $p=0$ case of our twisted model (TM)
\cite{cgm2} well accounts for the symmetries of the model under study. In
such a framework we give the whole primary fields content of the theory on
the plane and exhibit the ground state wave function.

In Section 4 the symmetry properties of the ground state conformal blocks
are analyzed and its relation with their topological properties shown.

In Section 5, starting from our CFT results, we show that the ground state
is degenerate, the different states being accessible by adiabatic flux
change techniques. Such a degeneracy is shown to be strictly related to the
presence in the spectrum of quasiparticles with non abelian statistics and
can be lifted non perturbatively through vortices tunneling.

In Section 6 some comments and outlooks are given.

In the Appendix we recall briefly the boundary states introduced in Ref.
\cite{noi1} in the framework of our TM.

\section{Josephson junctions ladders with Mobius boundary conditions}

In this Section we briefly describe the system we will study in the
following, that is a closed ladder of Josephson junctions (see Fig.1) with
Mobius boundary conditions. With each site $i$ we associate a phase $\varphi
_{i}$ and a charge $2en_{i}$, representing a superconducting grain coupled
to its neighbours by Josephson couplings; $n_{i}$ and $\varphi _{i}$ are
conjugate variables satisfying the usual phase-number commutation relation.
The system is described by the quantum phase model (QPM) Hamiltonian:
\begin{equation}
H=-\frac{E_{C}}{2}\sum_{i}\left( \frac{\partial }{\partial \varphi _{i}}%
\right) ^{2}-\sum_{\left\langle ij\right\rangle }E_{ij}\cos \left( \varphi
_{i}-\varphi _{j}-A_{ij}\right) ,  \label{act0}
\end{equation}
where $E_{C}=\frac{\left( 2e\right) ^{2}}{C}$ ($C$ being the grain capacitance) is
the charging energy at site $i$, while the second term is the Josephson
coupling energy between sites $i$ and $j$ and the sum is over nearest
neighbours. $A_{ij}=\frac{2\pi }{\Phi _{0}}$ $\int_{i}^{j}A{\cdot }dl$ is
the line integral of the vector potential associated to an external magnetic
field $B$ and $\Phi _{0}=\frac{hc}{2e}$ is the superconducting flux quantum.
The gauge invariant sum around a plaquette is $\sum_{p}A_{ij}=2\pi f$ with $%
f=\frac{\Phi }{\Phi _{0}}$, where $\Phi $ is the flux threading each
plaquette of the ladder.

\begin{figure}[tbp]
\centering\includegraphics*[width=0.6\linewidth]{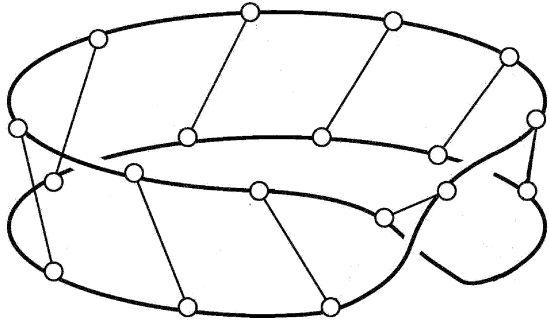}
\caption{Josephson junction ladder with Mobius boundary conditions}
\label{figura1}
\end{figure}

Let us label the phase fields on the two legs with $\varphi _{i}^{\left(
a\right) }$, $a=1,2$ and assume $E_{ij}=E_{x}$ for horizontal links and $%
E_{ij}=E_{y}$ for vertical ones. Let us also make the gauge choice $%
A_{ij}=+\pi f$ for the upper links, $A_{ij}=-\pi f$ for the lower ones and $%
A_{ij}=0$ for the vertical ones, which corresponds to a vector potential
parallel to the ladder and taking opposite values on upper and lower
branches.

Thus the effective quantum Hamiltonian (\ref{act0}) can be written as \cite
{ladder}:
\begin{eqnarray}
-H &=&\frac{E_{C}}{2}\sum_{i}\left[ \left( \frac{\partial }{\partial \varphi
_{i}^{\left( 1\right) }}\right) ^{2}+\left( \frac{\partial }{\partial
\varphi _{i}^{\left( 2\right) }}\right) ^{2}\right] +  \notag \\
&&\sum_{i}\left[ E_{x}\sum_{a=1,2}\cos \left( \varphi _{i+1}^{\left(
a\right) }-\varphi _{i}^{\left( a\right) }+\left( -1\right) ^{a}\pi f\right)
+E_{y}\cos \left( \varphi _{i}^{\left( 1\right) }-\varphi _{i}^{\left(
2\right) }\right) \right] .  \label{ha1}
\end{eqnarray}

The correspondence between the effective quantum Hamiltonian (\ref{ha1}) and
our TM model can be best traced performing the change of variables \cite
{ladder}: $\varphi _{i}^{\left( 1\right) }=X_{i}+\phi _{i}$, $\varphi
_{i}^{\left( 2\right) }=X_{i}-\phi _{i}$, so getting:
\begin{eqnarray}
H &=&-\frac{E_{C}}{2}\sum_{i}\left[ \left( \frac{\partial }{\partial X_{i}}%
\right) ^{2}+\left( \frac{\partial }{\partial \phi _{i}}\right) ^{2}\right] -
\notag \\
&&\sum_{i}\left[ 2E_{x}\cos \left( X_{i+1}-X_{i}\right) \cos \left( \phi
_{i+1}-\phi _{i}-\pi f\right) +E_{y}\cos \left( 2\phi _{i}\right) \right]
\label{ha2}
\end{eqnarray}
where $X_{i}$, $\phi _{i}$ (i.e. $\varphi _{i}^{\left( 1\right) }$, $\varphi
_{i}^{\left( 2\right) }$) are only phase deviations of each order parameter
from the commensurate phase and should not be identified with the phases of
the superconducting grains \cite{ladder}.

When $f=\frac{1}{2}$ and $E_{C}=0$ (classical limit) the ground state of the
$1D$ frustrated quantum XY (FQXY) model displays - in addition to the
continuous $U(1)$ symmetry of the phase variables - a discrete $Z_{2}$
symmetry associated with an antiferromagnetic pattern of plaquette
chiralities $\chi _{p}=\pm 1$, measuring the two opposite directions of the
supercurrent circulating in each plaquette. The evidence for a chiral phase
in Josephson junction ladders has been investigated in Ref. \cite{nishi}
while a field theoretical description of chiral order is developed in \cite
{azaria}.

Performing the continuum limit of the Hamiltonian (\ref{ha2}):
\begin{eqnarray}
-H &=&\frac{E_{C}}{2}\int dx\left[ \left( \frac{\partial }{\partial X}%
\right) ^{2}+\left( \frac{\partial }{\partial \phi }\right) ^{2}\right] +
\notag \\
&&\int dx\left[ E_{x}\left( \frac{\partial X}{\partial x}\right)
^{2}+E_{x}\left( \frac{\partial \phi }{\partial x}-\frac{\pi }{2}\right)
^{2}+E_{y}\cos \left( 2\phi \right) \right]  \label{ha3}
\end{eqnarray}
we see that the $X$ and $\phi $ fields are decoupled. In fact the $X$ term
of the above Hamiltonian is that of a free quantum field theory while the $%
\phi $ one coincides with the quantum sine-Gordon model. Through an
imaginary-time path-integral formulation of such a model \cite{zinn} it can
be shown that the $1D$ quantum problem maps into a $2D$ classical
statistical mechanics system, the $2D$ fully frustrated XY model, where the
parameter $\alpha =\left( \frac{E_{x}}{E_{C}}\right) ^{\frac{1}{2}}$ plays
the role of an inverse temperature \cite{ladder}. For small $E_{C}$ there is
a gap for creation of kinks in the antiferromagnetic pattern of $\chi _{p}$
and the ground state has quasi long range chiral order. We work in the
regime ${E_{x}}\gg {E_{y}}$ where the ladder is well described by a CFT with
central charge $c=2$.

We are now ready to introduce the modified ladder \cite{noi}, see Fig. 1. In
order to do so let us first require the $\varphi ^{\left( a\right) }$, $%
a=1,2 $, variables to recover the angular nature by compactification of both
the up and down fields. In such a way the XY-vortices, causing the
Kosterlitz-Thouless transition, are recovered. As a second step let us
introduce at point $x=0$ a defect which couples the up and down edges
through its interaction with the two legs, that is let us close the ladder
and impose Mobius boundary conditions. In the limit of strong coupling such
an interaction gives rise to non trivial boundary conditions for the fields
\cite{noi1}. In the following we give further details on such an issue, in
particular we adopt the $m$-reduction technique \cite{VM}\cite{cgm2}, which
accounts for non trivial boundary conditions \cite{noi1} for the Josephson
ladder in the presence of a defect line. In the Appendix the relevant chiral
fields ${{\varphi}_e} ^{\left( a\right) }$, $a=1,2$, which emerge from such
conditions, are explicitly constructed, by using the folding procedure.

\section{$m$-reduction technique}

In this Section we focus on the $m$-reduction technique for the special $m=2$
case and apply it to the system described by the Hamiltonian (\ref{ha3}). In
the Appendix each phase field $\varphi ^{\left( a\right) }$ is written as a
sum of two fields of opposite chirality defined on an half-line, because of
the presence of a defect at $x=0$. Within a ''bosonization'' framework it is
shown there how it is possible to reduce to a problem with two chiral fields
$\varphi _{e}^{\left( a\right) }$, $a=1,2$, each defined on the whole $x-$%
axis, and the corresponding dual fields. Now we identify in the continuum
such chiral phase fields $\varphi _{e}^{\left( a\right) }$, $a=1,2$, each
defined on the corresponding leg, with the two chiral fields $Q^{\left(
a\right) }$, $a=1,2$ of our CFT, the TM, with central charge $c=2$.

In order to construct such fields we start from a CFT with $c=1$ described
in terms of a scalar chiral field $Q$ compactified on a circle with radius $%
R^{2}=2$, explicitly given by:
\begin{equation}
Q(z)=q-i\,p\,lnz+i\sum_{n\neq 0}\frac{a_{n}}{n}z^{-n}  \label{modes}
\end{equation}
with $a_{n}$, $q$ and $p$ satisfying the commutation relations $\left[
a_{n},a_{n^{\prime }}\right] =n\delta _{n,n^{\prime }}$ and $\left[ q,p%
\right] =i$; its primary fields are the vertex operators $U^{\alpha_{l} }\left(
z\right) =:e^{i\alpha_{l} Q(z)}:$ and $\alpha_{l}=\frac{l}{\sqrt{2}}, l=1,2$. It is possible to give a plasma description
through the relation $\left| \psi \right| ^{2}=e^{-\beta H_{eff}}$ where $%
\psi \left( z_{1},...,z_{N}\right) =\left\langle N\alpha
|\prod_{i=1}^{N}U^{\alpha_{l} }(z_{i})|0\right\rangle =\prod_{i<j=1}^{N}\left(
z_{i}-z_{j}\right) ^{\frac{l^{2}}{2}}$ is the ground state wave function. It can be
immediately seen that $H_{eff}=-l^{2}\sum_{i<j=1}^{N}\ln \left|
z_{i}-z_{j}\right| $ and $\beta =\frac{2}{R^{2}}=1$, that is only vorticity $%
v=1,2$ vortices are present in the plasma.

Starting from such a CFT mother theory one can use the $m$-reduction
procedure, which consists in considering the subalgebra generated only by
the modes in eq. (\ref{modes}) which are a multiple of an integer $m$, so
getting a $c=m$ orbifold CFT (daughter theory, i.e. the twisted model (TM))
\cite{cgm2}. With respect to the special $m=2$ case, the fields in the
mother CFT can be organized into components which have well defined
transformation properties under the discrete $Z_{2}$ (twist) group, which is
a symmetry of the TM. By using the mapping $z\rightarrow z^{1/2}$ and by
making the identifications $a_{2n+l}\longrightarrow \sqrt{2}a_{n+l/2}$, $%
q\longrightarrow \frac{1}{\sqrt{2}}q$ the $c=2$ daughter CFT is obtained. It
is interesting to notice that such a daughter CFT gives rise to a vortices
plasma of half integer vorticity, that is to a fully frustrated XY model, as
it will appear in the following.

Its primary fields content can be expressed in terms of a $Z_{2}$-invariant
scalar field $X(z)$, given by
\begin{equation}
X(z)=\frac{1}{2}\left( Q^{(1)}(z)+Q^{(2)}(z)\right) ,  \label{X}
\end{equation}
describing the continuous phase sector of the new theory, and a twisted
field
\begin{equation}
\phi (z)=\frac{1}{2}\left( Q^{(1)}(z)-Q^{(2)}(z)\right) ,  \label{phi}
\end{equation}
which satisfies the twisted boundary conditions $\phi (e^{i\pi }z)=-\phi (z)$
\cite{cgm2}. Such fields coincide with the ones introduced in eq. (\ref{ha3}%
).

The whole TM theory decomposes into a tensor product of two CFTs, a twisted
invariant one with $c=\frac{3}{2}$ and the remaining $c=\frac{1}{2}$ one
realized by a Majorana fermion in the twisted sector. In the $c=\frac{3}{2}$
subtheory the primary fields are composite vertex operators $V\left(
z\right) =U_{X}^{\alpha_{l}}\left( z\right) \psi \left( z\right) $ or $V_{qh}\left(
z\right) =U_{X}^{\alpha_{l}}\left( z\right) \sigma \left( z\right) $, where $U_{X}^{\alpha_{l}}\left(
z\right) =\frac{1}{\sqrt{z}}:e^{i\alpha_{l} X(z)}:$ is the vertex of the
charged\ sector with $\alpha_{l}=\frac{l}{2}, l=1,...,4$ for the $SU(2)$ Cooper pairing symmetry
used here.

Regarding the other\ component, the highest weight state in the neutral
sector can be classified by the two chiral operators:
\begin{eqnarray}
\psi \left( z\right) &=&\frac{1}{2\sqrt{z}}\left( :e^{i\sqrt{2} \phi \left(
z\right) }:+:e^{i\sqrt{2} \phi \left( -z\right) }:\right) ,~~~~~~  \notag \\
\overline{\psi }\left( z\right) &=&\frac{1}{2\sqrt{z}}\left( :e^{i\sqrt{2}
\phi \left( z\right) }:-:e^{i\sqrt{2} \phi \left( -z\right) }:\right) ;
\label{neu11}
\end{eqnarray}
which correspond to two $c=\frac{1}{2}$ Majorana fermions with Ramond
(invariant under the $Z_{2}$ twist) or Neveu-Schwartz ($Z_{2}$ twisted)
boundary conditions \cite{cgm2} in a fermionized version of the theory. Let
us point out that the energy-momentum tensor of the Ramond part of the
neutral sector develops a cosine term:
\begin{equation}
T_{\psi }\left( z\right) =-\frac{1}{4}\left( \partial \phi \right) ^{2}-%
\frac{1}{16z^{2}}\cos \left( 2\sqrt{2}\phi \right) ,  \label{tn1}
\end{equation}
a clear signature of a tunneling phenomenon which selects a new stable
vacuum, the linear superposition of the two ground states. The Ramond fields
are the degrees of freedom which survive after the tunneling and the $Z_{2}$
(orbifold) symmetry, which exchanges the two Ising fermions, is broken.

So the whole energy-momentum tensor within the $c=\frac{3}{2}$ subtheory is:
\begin{equation}
T=T_{X}\left( z\right) +T_{\psi }\left( z\right) =-\frac{1}{2}\left(
\partial X\right) ^{2}-\frac{1}{4}\left( \partial \phi \right) ^{2}-\frac{1}{%
16z^{2}}\cos \left( 2\sqrt{2}\phi \right) .  \label{ttt1}
\end{equation}
The correspondence with the Hamiltonian in eq. (\ref{ha3}) is more evident
once we observe that the neutral current $\partial \phi $ appearing above
coincides with the term $\partial \phi -\frac{\pi }{2}$ of eq. (\ref{ha3}),
since the $\frac{\pi }{2}$-term coming there from the frustration condition,
here it appears in $\partial \phi $ as a zero mode, i.e. a classical mode.
Besides the fields appearing in eq. (\ref{neu11}) there are the $\sigma
\left( z\right) $ fields, also called the twist fields, which appear in the
primary fields $V_{qh}\left( z\right) $ combined to a vertex with charge $%
\frac{e}{4}$. The twist fields have non local properties and decide also for
the non trivial properties of the vacuum state, which in fact can be twisted
or not in our formalism. Such a property for the vacuum is more evident for
the torus topology, where the $\sigma $-field is described by the conformal
block $\chi _{\frac{1}{16}}$ (see Section 4).

The evidence of a phase transition in ladder systems at $c={\frac{3}{2}}$
has been investigated in \cite{gritsev} within a CFT framework. Within this
framework the ground state wave function is described as a correlator of $%
N_{2e}$ Cooper pairs:
\begin{equation}
<N_{2e}\alpha |\prod_{i=1}^{N_{2e}}V^{\sqrt{2}}(z_{i})|0>=\prod_{i<i^{\prime
}=1}^{N_{2e}}(z_{i}-z_{i^{\prime }})Pf\left( \frac{1}{z_{i}-z_{i^{\prime }}}%
\right)  \label{pff}
\end{equation}
where $Pf\left( \frac{1}{z_{i}-z_{i^{\prime }}}\right) =\mathcal{A}\left(
\frac{1}{z_{1}-z_{2}}\frac{1}{z_{3}-z_{4}}\dots \right) $ is the
antisymmetrized product over pairs of Cooper pairs, so reproducing well
known results \cite{MR}. In a similar way we also are able to evaluate
correlators of $N_{2e}$ Cooper pairs in the presence of (quasi-hole)
excitations \cite{MR}\cite{cgm2} with non Abelian statistics \cite{nayak}. It
is now interesting to notice that the charged contribution appearing in the
correlator of $N_{e}$ electrons is just: $<N_{e}\alpha
|\prod_{i=1}^{N_{2e}}U_{X}^{1/2}(z_{i})|0>=\prod_{i<i^{\prime
}=1}^{N_{e}}(z_{i}-z_{i^{\prime }})^{1/4}$, giving rise to a vortices plasma
with $H_{eff}=-\frac{1}{4}\sum_{i<j=1}^{N}\ln \left\vert
z_{i}-z_{j}\right\vert $ at the corresponding "temperature" $\beta =\frac{2}{R_{X}^{2}}=2$, that
is it describes vortices with vorticity $v=\frac{1}{2}$!

On closed annulus geometries, as it is the discretized analogue of a torus,
we must properly account for boundary conditions at the ends of the finite
lattice since they determine in the continuum the pertinent conformal blocks
yielding the statistics of quasiparticles as well as the ground state
degeneracy. We have two possible boundary conditions which correspond to two
different ways to close the double lattice, i.e. $\varphi _{N}^{\left(
a\right) }\rightarrow \varphi _{1}^{\left( a\right) }$ or $\varphi
_{N}^{\left( a+1\right) }\rightarrow \varphi _{1}^{\left( a\right) }$, $%
a=1,2 $. It is not difficult to work out that, for the ladder case, the
twisted boundary conditions can be implemented only on the odd lattice.

\begin{figure}[tbp]
\centering\includegraphics*[width=0.9\linewidth]{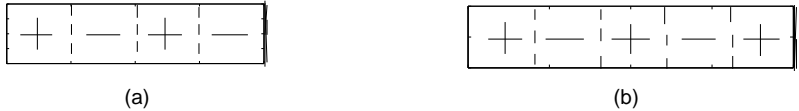}
\caption{boundary conditions: (a) untwisted; (b) twisted. }
\label{figura2}
\end{figure}

Indeed for $f=\frac{1}{2}$ the ladder is invariant under the shift of two
sites, so that there are two topologically inequivalent boundary conditions
for even or odd number of sites. In the even case the end sites are of the
same kind of the starting one, while in the odd case a ferromagnetic line
corresponds to an antiferromagnetic one. The even (odd) case corresponds to
untwisted (twisted) boundary conditions, as depicted in Fig. 2. The odd case
selects out two degenerate ground states which belong to a non trivial topological
sector, indeed the system may develop topological order in the twisted sector
of our theory. Let us notice that in the discrete case not all the vacua in
the different sectors of our theory are connected. For instance the states
in the untwisted sector, which correspond to the even ladder, are physically
disconnected from those of the twisted one, which correspond to the odd
ladder.

\section{Symmetry properties of the TM conformal blocks}

In Section 3 we identified our chiral fields $Q^{(a)}$ with the continuum
limit of the Josephson phase $\varphi ^{\left( a\right) }$ defined on the
two legs of the ladder respectively and considered non trivial boundary
conditions at its ends, so constructing a version in the continuum of the
discrete system. By using standard conformal field theory techniques it is
now possible to generate the torus topology, starting from the edge theory,
just defined in the previous Section. That is realized by evaluating the $N-$%
vertices correlator

\begin{equation}
\left\langle n\right| V_{\alpha }\left( z_{1}\right) ...V_{\alpha }\left(
z_{N}\right) e^{2\pi i\tau L_{0}}\left| n\right\rangle
\end{equation}
where $V_{\alpha }\left( z_{i}\right) $ is the generic primary field of
Section 3 representing the excitation at $z_{i}$, $L_{0}$ is the Virasoro
generator for dilatations and $\tau $ the proper time. The neutrality
condition $\sum \alpha =0$ must be satisfied and the sum over the complete
set of states $\left| n\right\rangle $ is indicating that a trace must be
taken. For the present paper it is not necessary to go through such a
calculation but it is very illuminating for the non expert reader to
pictorially represent the above operation in terms of an edge state (that is
a primary state defined at a given $\tau $) which propagates interacting
with external fields at $z_{1}...z_{N}$ and finally getting back to itself.
In such a way a $2D$ surface is generated with the torus topology. From such a 
picture it is evident then how the degeneracy of the non perturbative ground state
is closely related to the number of primary states.
Furthermore, since in this paper we are interested in the understanding of
the topological properties of the system, we can consider only the center of
mass contribution in the above correlator, so neglecting its short distances
properties. To such an extent the one-point functions are extensively
reported in the following.

On the torus \cite{cgm4} the TM primary fields are described in terms of the
conformal blocks of the $Z_{2}$-invariant $c=3/2$ subtheory and of the non
invariant $c=1/2$ Ising model, so reflecting the decomposition on the plane
outlined in the previous Section. The following characters 
\begin{align*}
\bar{\chi}_{0}(w_{n}|\tau )& =\frac{1}{2}\left( \sqrt{\frac{\theta
_{3}(w_{n}|\tau )}{\eta (\tau )}}+\sqrt{\frac{\theta _{4}(w_{n}|\tau )}{\eta
(\tau )}}\right)  \\
\bar{\chi}_{\frac{1}{2}}(w_{n}|\tau )& =\frac{1}{2}\left( \sqrt{\frac{\theta
_{3}(w_{n}|\tau )}{\eta (\tau )}}-\sqrt{\frac{\theta _{4}(w_{n}|\tau )}{\eta
(\tau )}}\right)  \\
\bar{\chi}_{\frac{1}{16}}(w_{n}|\tau )& =\sqrt{\frac{\theta _{2}(w_{n}|\tau )%
}{2\eta (\tau )}}
\end{align*}
express the primary fields content of the Ising model with Neveu-Schwartz ($%
Z_{2}$ twisted)  boundary conditions, while 
\begin{eqnarray}
\chi _{(0)}^{c=3/2}(w_{n}|w_{c}|\tau ) &=&\chi _{0}(w_{n}|\tau
)K_{0}(w_{c}|\tau )+\chi _{\frac{1}{2}}(w_{n}|\tau )K_{2}(w_{c}|\tau )
\label{mr1} \\
\chi _{(1)}^{c=3/2}(w_{n}|w_{c}|\tau ) &=&\chi _{\frac{1}{16}}(w_{n}|\tau
)\left( K_{1}(w_{c}|\tau )+K_{3}(w_{c}|\tau )\right)   \label{mr2} \\
\chi _{(2)}^{c=3/2}(w_{n}|w_{c}|\tau ) &=&\chi _{\frac{1}{2}}(w_{n}|\tau
)K_{0}(w_{c}|\tau )+\chi _{0}(w_{n}|\tau )K_{2}(w_{c}|\tau )  \label{mr3}
\end{eqnarray}
represent those of the $Z_{2}$-invariant $c=3/2$ \ CFT. They are given in terms of a ``charged''\ %
$K_{\alpha }(w_{c}|\tau )$ contribution, (see definition given below) and a ``neutral''\ one $%
\chi _{\beta }(w_{n}|\tau )$, (the conformal blocks of the Ising Model), where $w_{c}=\frac{1}{2\pi i}\ln z_{c}$ and $w_{n}=\frac{1}{%
2\pi i}\ln z_{n}$ are the torus variables of ``charged''\ and ``neutral''\
components respectively.

In order to understand the physical significance of the $c=2$ conformal
blocks in terms of the charged low energy excitations of the system, let us
evidence their electric charge (magnetic flux contents in the dual theory,
which is obtained by exchanging the compactification radius $%
R_{e}^{2}\rightarrow R_{m}^{2}$ in the charged sector of the CFT). In order
to do so let us consider the ``charged''\ sector conformal blocks appearing
in eqs. (\ref{mr1}-\ref{mr3}): 
\begin{equation}
K_{2l+i}(w|\tau )=\frac{1}{\eta \left( \tau \right) }\Theta \left[ 
\begin{array}{c}
\frac{2l+i}{4} \\ 
0
\end{array}
\right] (2w|4\tau ),~~~~~\forall \left( l,i\right) \in \left( 0,1\right)
^{2}.  \label{chp}
\end{equation}
They correspond to primary fields with conformal dimensions $h_{2l+i}=\frac{1%
}{2}\alpha _{\left( l,i\right) }^{2}=\frac{1}{2}\left( \frac{2l+i}{2}%
+2\delta _{\left( l+i\right) ,0}\right) ^{2}$ and electric charges $2e\left( 
\frac{\alpha _{\left( l,i\right) }}{R_{X}}\right) $(magnetic charges in the
dual theory $\frac{hc}{2e}\left( \alpha _{\left( l,i\right) }R_{X}\right) $%
), $R_{X}=1$ being the compactification radius. More explicitly the electric
charges (magnetic charges in the dual theory) are the following: 
\begin{equation}
\begin{array}{cccc}
l=0 & i=0 & q_{e}=4e & \left( q_{m}=2\frac{hc}{2e}\right)  \\ 
l=1 & i=0 & q_{e}=2e & \left( q_{m}=\frac{hc}{2e}\right)  \\ 
l=0 & i=1 & q_{e}=e & \left( q_{m}=\frac{1}{2}\frac{hc}{2e}\right)  \\ 
l=1 & i=1 & q_{e}=3e & \left( q_{m}=\frac{3}{2}\frac{hc}{2e}\right) 
\end{array}
.
\end{equation}
If we now turn to the whole $c=2$ theory, the characters of the twisted sector
are given by: 
\begin{eqnarray}
\chi _{(0)}^{+}(w_{n}|w_{c}|\tau ) &=&\bar{\chi}_{\frac{1}{16}}(w_{n}|\tau
)\left( \chi _{0}^{c=3/2}(w_{n}|w_{c}|\tau )+\chi
_{2}^{c=3/2}(w_{n}|w_{c}|\tau )\right)   \label{tw1} \\
&=&\bar{\chi}_{\frac{1}{16}}\left( \chi _{0}+\chi _{\frac{1}{2}}\right)
\left( K_{0}+K_{2}\right)   \notag \\
\chi _{(1)}^{+}(w_{n}|w_{c}|\tau ) &=&\left( \bar{\chi}_{0}(w_{n}|\tau )+%
\bar{\chi}_{\frac{1}{2}}(w_{n}|\tau )\right) \chi
_{1}^{c=3/2}(w_{n}|w_{c}|\tau )  \label{tw2} \\
&=&\chi _{\frac{1}{16}}\left( \bar{\chi}_{0}+\bar{\chi}_{\frac{1}{2}}\right)
\left( K_{1}+K_{3}\right)   \notag \\
\chi _{(0)}^{-}(w_{n}|w_{c}|\tau ) &=&\bar{\chi}_{\frac{1}{16}}(w_{n}|\tau
)\left( \chi _{0}^{c=3/2}(w_{n}|w_{c}|\tau )-\chi
_{2}^{c=3/2}(w_{n}|w_{c}|\tau )\right)  \\
&=&\bar{\chi}_{\frac{1}{16}}\left( \chi _{0}-\chi _{\frac{1}{2}}\right)
\left( K_{0}-K_{2}\right)   \notag \\
\chi _{(1)}^{-}(w_{N}|w_{c}|\tau ) &=&\left( \bar{\chi}_{0}(w_{n}|\tau )-%
\bar{\chi}_{\frac{1}{2}}(w_{n}|\tau )\right) \chi
_{1}^{c=3/2}(w_{n}|w_{c}|\tau ) \\
&=&\chi _{\frac{1}{16}}\left( \bar{\chi}_{0}-\bar{\chi}_{\frac{1}{2}}\right)
\left( K_{1}+K_{3}\right) .  \notag
\end{eqnarray}

Furthermore the characters of the untwisted sector are \cite{cgm4}: 
\begin{align}
\tilde{\chi}_{(0)}^{+}(w_{n}|w_{c}|\tau )& =\bar{\chi}_{0}(w_{n}|\tau )\chi
_{(0)}^{c=3/2}(w_{n}|w_{c}|\tau )+\bar{\chi}_{\frac{1}{2}}(w_{n}|\tau )\chi
_{(2)}^{c=3/2}(w_{n}|w_{c}|\tau )  \label{vac1} \\
& =\left( \bar{\chi}_{0}\chi _{0}+\bar{\chi}_{\frac{1}{2}}\chi _{\frac{1}{2}%
}\right) K_{0}+\left( \bar{\chi}_{0}\chi _{\frac{1}{2}}+\bar{\chi}_{\frac{1}{%
2}}\chi _{0}\right) K_{2}  \notag \\
\tilde{\chi}_{(1)}^{+}(w_{n}|w_{c}|\tau )& =\bar{\chi}_{0}(w_{n}|\tau )\chi
_{(2)}^{c=3/2}(w_{n}|w_{c}|\tau )+\bar{\chi}_{\frac{1}{2}}(w_{n}|\tau )\chi
_{(0)}^{c=3/2}(w_{n}|w_{c}|\tau )  \label{vac2} \\
& =\left( \bar{\chi}_{0}\chi _{\frac{1}{2}}+\bar{\chi}_{\frac{1}{2}}\chi
_{0}\right) K_{0}+\left( \bar{\chi}_{0}\chi _{0}+\bar{\chi}_{\frac{1}{2}%
}\chi _{\frac{1}{2}}\right) K_{2}  \notag \\
\tilde{\chi}_{(0)}^{-}(w_{n}|w_{c}|\tau )& =\bar{\chi}_{0}(w_{n}|\tau )\chi
_{(0)}^{c=3/2}(w_{n}|w_{c}|\tau )-\bar{\chi}_{\frac{1}{2}}(w_{n}|\tau )\chi
_{(2)}^{c=3/2}(w_{n}|w_{c}|\tau )  \label{vac3} \\
& =\left( \bar{\chi}_{0}\chi _{0}-\bar{\chi}_{\frac{1}{2}}\chi _{\frac{1}{2}%
}\right) K_{0}+\left( \bar{\chi}_{0}\chi _{\frac{1}{2}}-\bar{\chi}_{\frac{1}{%
2}}\chi _{0}\right) K_{2}  \notag \\
\tilde{\chi}_{(1)}^{-}(w_{n}|w_{c}|\tau )& =\bar{\chi}_{0}(w_{n}|\tau )\chi
_{(2)}^{c=3/2}(w_{n}|w_{c}|\tau )-\bar{\chi}_{\frac{1}{2}}(w_{n}|\tau )\chi
_{(0)}^{c=3/2}(w_{n}|w_{c}|\tau )  \label{vac4} \\
& =\left( \bar{\chi}_{0}\chi _{\frac{1}{2}}-\bar{\chi}_{\frac{1}{2}}\chi
_{0}\right) K_{0}+\left( \bar{\chi}_{0}\chi _{0}-\bar{\chi}_{\frac{1}{2}%
}\chi _{\frac{1}{2}}\right) K_{2}  \notag \\
\tilde{\chi}_{(0)}(w_{n}|w_{c}|\tau )& =\bar{\chi}_{\frac{1}{16}}(w_{n}|\tau
)\chi _{(1)}^{c=3/2}(w_{n}|w_{c}|\tau )=\bar{\chi}_{\frac{1}{16}}\chi _{%
\frac{1}{16}}\left( K_{1}+K_{3}\right) .  \label{vac5}
\end{align}
Such a factorization is a consequence of the parity selection rule ($m$%
-ality), which gives a gluing condition for the ``charged''\ and
``neutral''\ excitations. The conformal blocks given above represent the
collective states of highly correlated vortices, which appear to be
incompressible. In order to show the corresponding symmetry properties it is
useful to give a pictorial description of the conformal blocks appearing in
eq. (\ref{chp}). To such an extent let us imagine to cut the torus along the 
$A$-cycle. The different primary fields then can be seen as excitations
which propagate along the $B$-cycle and interact with the external Cooper
pair at point $w=\frac{1}{2\pi i}\ln z$. We can now test the symmetry
properties of the characters of the theory (given above) by simply
evaluating the Bohm-Aharonov phase they pick up while a Cooper pair is taken
along the closed $A$-cycle. In order to do that, it is important to notice
that the transport of the ''Cooper pair'' from the upper (with isospin up)
leg to the down (with isospin down) leg can be realized by a translation of
the variables $w_{c}$ and $w_{n}$, which must be identical for the
''charged'' and the ''neutral'' sectors. So, under a $2\pi $-rotation the
torus variables transforms as $\Delta w_{c}=\Delta w_{n}=1$. It is easy to
check that: 
\begin{equation}
K_{0,2}(w_{c}+1|\tau )=K_{0,2}(w_{c}|\tau ),~~K_{1,3}(w_{c}+1|\tau
)=-K_{1,3}(w_{c}|\tau ).  \label{c18}
\end{equation}
The change in sign given in eq. (\ref{c18}) under a $2\pi $-rotation is
strictly related to the presence in the spectrum of excitations carrying
fractionalized charge quanta. Now, turning on also the neutral sector
contribution in the Cooper pair transport along the $A$-cycle we obtain in a
straightforward way: 
\begin{equation}
\chi _{0,\frac{1}{2}}(w_{n}+1|\tau )=\chi _{0,\frac{1}{2}}(w_{n}|\tau
),~~\chi _{\frac{1}{16}}(w_{n}+1|\tau )=i\chi _{\frac{1}{16}}(w_{n}|\tau )
\label{c19}
\end{equation}
the same is true for the characters $\bar{\chi}_{\beta }$. A phase $i=e^{i%
\frac{\pi }{2}}$ appears in the $\chi _{\frac{1}{16}}$ character due to the
presence of a half-flux.

As a result the ground state described by eq. (\ref{vac5}): 
\begin{equation}
\tilde{\chi}_{(0)}(w_{n}|w_{c}|\tau )=\bar{\chi}_{\frac{1}{16}}\chi _{\frac{1%
}{16}}\left( K_{1}+K_{3}\right)  \label{pt2}
\end{equation}
does not change sign under, $\Delta w_{c}=\Delta w_{n}=1$, the transport of
a Cooper pair along the closed $A$-cycle. In fact the negative sign coming
from the continuous phase sector being compensated by the negative sign
coming from the other sector! Of course the same is true for all the other
characters of the untwisted sector, i.e. we cannot trap a half flux quantum
in the hole in the untwisted sector.

Instead in the twisted sector the ground state wave-functions show a non
trivial behavior. In fact under $\Delta w_{c}=\Delta w_{n}=1$%
\begin{equation}
\chi _{\left( 0\right) }^{\pm }(w_{n}+1|w_{c}+1|\tau )=+i\chi _{\left(
0\right) }^{\pm }(w_{n}|w_{c}|\tau ),~~\chi _{\left( 1\right) }^{\pm
}(w_{n}+1|w_{c}+1|\tau )=-i\chi _{\left( 1\right) }^{\pm }(w_{n}|w_{c}|\tau
),
\end{equation}
The change in phase given above evidences the presence of a half flux
quantum in the hole. In fact in the twisted case geometry (see Fig2) the
Cooper pair flows along the ladder and changes isospin in a $2\pi $-period,
so implying that in such a case the transport of a Cooper pair from
a given point $w$ on the $A$-cycle to the same point has a $4\pi $-period,
that is it corresponds to $\Delta w_{c}=\Delta w_{n}=2$. Under this
transformation the characters get the following non trivial phase: 
\begin{equation}
\chi _{\left( 0,1\right) }^{\pm }(w_{n}+2|w_{c}+2|\tau )=-\chi _{\left(
0,1\right) }^{\pm }(w_{n}|w_{c}|\tau ),  \label{half-flux-trapped}
\end{equation}
so explicitly evidencing the trapping of $\frac{1}{2}\frac{hc}{2e}$ in the
hole.

It is worthwhile to notice that the properties just discussed are
independent of the short distance properties of the vortices plasma, the
only crucial requirement for its stability being the neutrality condition.

\section{Topological order and ``protected'' qubits}

The aim of this Section is to fully exploit the issue of topological order
in a quantum JJL. In order to meet such a request let us use the results of
the $2$-reduction technique for the torus topology \cite{cgm4}. For closed
geometries the JJL with Mobius boundary conditions gives rise to a line
defect in the bulk. So it becomes mandatory to use a folding procedure to
map the problem with a defect line into a boundary one, where the defect
line appears as a boundary state. The TM boundary states have been
constructed in Ref. \cite{noi1} together with the corresponding chiral
partition functions and briefly recalled in the Appendix. In particular we
get an ''untwisted''\ sector and a ''twisted''\ one, corresponding to
periodic and Mobius boundary conditions respectively. That gives rise to an
essential difference in the low energy spectrum for the system under study:
in the first case only two-spinon excitations are possible (i.e. ${SU(2)}_{2}
$ integer spin representations) while, in the last case, the presence of a
topological defect provides a clear evidence of single-spinon excitations
(i.e. ${SU(2)}_{2}$ half-integer spin representations). All that takes place
in close analogy with spin-1/2 closed zigzag ladders, which are expected to
map to fully frustrated Josephson ladders in the extremely quantum limit. In
this way all previous results obtained by means of exact diagonalization
techniques \cite{mobius} are recovered.

In the following we discuss topological order referring to the characters
which in turn are related to the different boundary states present in the
system through such chiral partition functions \cite{noi}. We can build a
topological invariant $\mathcal{P}=\prod\limits_{\gamma }\chi _{p}$ where $%
\gamma $ is a closed contour that goes around the hole. Such a choice allows
us to define two degenerate ground states with $\mathcal{P}=1$ and $\mathcal{%
P}=-1$ respectively, labelled $\left\vert 0\right\rangle ,\left\vert
1\right\rangle $, as in Fig. 3(a): there must be an odd number of plaquettes
(odd ladder) to satisfy this condition which is imposed by the request of
topological protection. The value $\mathcal{P}=-1$ selects twisted boundary
conditions at the ends of the chain while $\mathcal{P}=1$ selects periodic
boundary conditions (see Fig. 3(b)). The case of finite discrete systems has
been discussed in detail in Ref. \cite{grimm}, our theory is its counterpart
in the continuum.

\begin{figure}[tbp]
\centering\includegraphics*[width=0.7\linewidth]{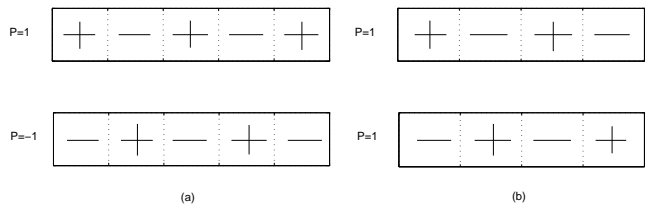}
\caption{degenerate ground states for: (a) odd ladder, $\mathcal{P}=\pm 1$;
(b) even ladder, $\mathcal{P}=1$. }
\label{figura3}
\end{figure}

In fact it is now possible to identify the two degenerate ground states shown in
Fig.3(a) with the twisted characters (\ref{tw1}-\ref{tw2}): 
\begin{equation}
\left| 0\right\rangle \rightarrow \chi _{\left( 0\right) }^{+}(w_{c}|\tau
),~~\left| 1\right\rangle \rightarrow \chi _{\left( 1\right)
}^{+}(w_{c}|\tau ),  \label{gsd}
\end{equation}
where it is meant that $w_{n}$ is fixed to zero (see also (\ref{p6}) and (%
\ref{p7}) of Appendix), then it is possible to remove such a degeneracy
through vortices tunneling. The last operation is also needed in order to
prepare the qubit in a definite state.

In fact such two states can be distinguished from all the other ones. First
of all, being in the twisted sector they trap a half flux quantum inside the
central hole (see (\ref{half-flux-trapped})). The trapped half flux quantum
can be experimentally detected, so giving a way to read out the twisted
states of the system.

Now, upon performing an adiabatic change of local magnetic fields which
drags one half vortex across the system, i.e. through the transport of a
half flux quantum around the $B$-cycle of the torus, we are able to fully
distinguish such two states with respect to the other pair of states, $\chi
_{\left( 0\right) }^{-}$ and $\chi _{\left( 1\right) }^{-}$, of the twisted
sector.

In fact, the transformations 
\begin{eqnarray}
\chi _{(0)}^{+}(\tau /2|w_{C}+\tau /2|\tau ) &=&-ie^{-i\pi
w_{C}}q^{-3/8}\chi _{(1)}^{+}(w_{C}|\tau ) \\
\chi _{(1)}^{+}(\tau /2|w_{C}+\tau /2|\tau ) &=&ie^{-i\pi w_{C}}q^{-5/8}\chi
_{(0)}^{+}(w_{C}|\tau ),
\end{eqnarray}
\begin{eqnarray}
\chi _{(0)}^{-}(\tau /2|w_{C}+\tau /2|\tau ) &=&0 \\
\chi _{(1)}^{-}(\tau /2|w_{C}+\tau /2|\tau ) &=&0
\end{eqnarray}
just imply a flip among the two ground states of the system 
\begin{equation}
\left| 0\right\rangle \rightarrow \left| 1\right\rangle ~\text{and}~\left|
1\right\rangle \rightarrow \left| 0\right\rangle ,
\end{equation}
while the other two twisted states are annihilated.

Finally it should be noticed that the presence of the twist operators $%
\sigma $, described here by the Ising character $\chi _{\frac{1}{16}}$,
gives rise to non Abelian statistics, which can be evidenced by their fusion
rules $\sigma \sigma =\mathbf{1}+\psi $ \cite{MR}\cite{cgm4}.

\section{Brief summary with comments}

In this contribution we presented a simple collective description of a
ladder of Josephson junctions with a macroscopic half flux quanta trapped in
the hole of a ladder. It was shown how the phenomenon of flux
fractionalization takes place within the context of a $2D$ conformal field
theory with a $Z_{2}$ twist, the TM. The presence of a $Z_{2}$ symmetry
indeed accounts for more general boundary conditions for the fields
describing the Cooper pairs propagating on the ladder legs, which arise from
the presence of a magnetic impurity strongly coupled with the Josephson
phases. It should be noticed that it would be useful to extend our approach
to a generic frustration $f=\frac{1}{m}$. For closed geometries and in the
limit of the continuum the phase fields $\varphi ^{(a)}$ defined on the two
legs were identified with the two chiral Fubini fields $Q^{(a)}$ of our TM,
and a correspondence between the energy momentum density tensor for such
fields (or better the $X$ and $\phi $ fields of eqs. (\ref{X})-(\ref{phi}))
and the Hamiltonian of eq. (\ref{ha3}) traced. For such geometries it was
also indicated that the Kosterlitz-Thouless vortices were recovered.

Furthermore it was shown that for closed geometries the JJL with an impurity
gives rise to a line defect, which can be turned into a boundary state after
employing the folding procedure. That enabled us to derive the low energy
charged excitations of the system as provided by our description, with the
superconducting phase characterized by condensation of $4e$ charges and
gapped $2e$ excitations. Finally, by simply evaluating a Bohm-Aharonov
phase, it has been evidenced that non trivial symmetry properties for the
conformal blocks emerge due to the presence in the spectrum of
fractionalized flux quanta $\frac{1}{2}\left( \frac{hc}{2e}\right) $. As it
has been explained before, that signals the presence of a topological defect
in the twisted sector of the TM. The question of an emerging topological
order in the ground state together with the possibility of providing
protected states for the implementation of a solid state qubit has been also
addressed \cite{noi}. Notice also the different behavior under transport of
the $2e$ and $4e$ excitations as it is well evidenced by the Bohm-Aharonov phase.
Indeed while the transport of a $2e\ $ excitation along the $A$-cycle induces a $-1$ phase
factor, in the $4e$ excitation transport the phase factor is trivial \cite
{ioffe}. That is a consequence of the symmetry of the $4e$ field with respect
to the leg index.

Finally we have shown that Josephson junctions ladders with non trivial
geometry may develop topological order allowing for the implementation of
``protected''\ qubits, a first step toward the realization of an ideal solid
state quantum computer. Josephson junctions ladders with annular geometry
have been fabricated within the trilayer $Nb/Al-AlO_{x}/Nb$ technology and
experimentally investigated \cite{ustinov}. So in principle it could be
simple to conceive an experimental setup in order to test our predictions.

It is interesting to notice that the presence of a topological defect has
been experimentally evidenced very recently for a two layers quantum Hall
system, by measuring the conduction properties between the two edges of the
system \cite{deviatov}.

\section*{Appendix}

Here we recall briefly the TM boundary states (BS) recently
constructed in \cite{noi1}. For closed geometries, that is for the torus,
the JJL with an impurity gives rise to a line defect in the bulk. In a
theory with a defect line the interaction with the impurity gives rise to
the following non trivial boundary conditions for the fields:
\begin{equation}
\varphi _{L}^{\left( a\right) }\left( x=0\right) =\mp \varphi _{R}^{\left(
a\right) }\left( x=0\right) -\varphi _{0},\text{ \ \ }a=1,2.  \label{blr}
\end{equation}
In order to describe it we resort to the folding procedure. Such a procedure
is used in the literature to map a problem with a defect line (as a bulk
property) into a boundary one, where the defect line appears as a boundary
state of a theory which is not anymore chiral and its fields are defined in
a reduced region which is one half of the original one. Our approach, the
TM, is a chiral description of that, where the chiral $\phi $\ field defined
in ($-L/2$, $L/2)$ describes both the left moving component and the right
moving one defined in ($-L/2$, $\ 0$), ($0$, $L/2$) respectively, in the
folded description \cite{noi1}. Furthermore to make a connection with the TM
we consider more general gluing conditions:

$\phi _{L}(x=0)=\mp \phi _{R}(x=0)-\varphi _{0}$

the $-$($+$) sign staying for the twisted (untwisted) sector. We are then
allowed to use the boundary states given in \cite{Affleck} for the $c=1$
orbifold at the Ising$^{2}$ radius. The $X$ field, which is even under the
folding procedure, does not suffer any change in boundary conditions \cite
{noi1}. Let us now write each phase field as the sum $\varphi ^{\left(
a\right) }\left( x\right) =\varphi _{L}^{\left( a\right) }\left( x\right)
+\varphi _{R}^{\left( a\right) }\left( x\right) $ of left and right moving
fields defined on the half-line because of the defect located in $x=0$. Then
let us define for each leg the two chiral fields $\varphi _{e,o}^{\left(
a\right) }\left( x\right) =\varphi _{L}^{\left( a\right) }\left( x\right)
\pm \varphi _{R}^{\left( a\right) }\left( -x\right) $, each defined on the
whole $x-$axis \cite{boso}. In such a framework the dual fields $\varphi
_{o}^{\left( a\right) }\left( x\right) $ are fully decoupled because the
corresponding boundary interaction term in the Hamiltonian does not involve
them \cite{affleck}; they are involved in the definition of the conjugate
momenta $\Pi _{\left( a\right) }=\left( \partial _{x}\varphi _{o}^{\left(
a\right) }\right) =\left( \frac{\partial }{\partial \varphi _{e}^{\left(
a\right) }}\right) $ present in the quantum Hamiltonian. Performing the
change of variables $\varphi _{e}^{\left( 1\right) }=X+\phi $, $\varphi
_{e}^{\left( 2\right) }=X-\phi $ ($\varphi _{o}^{\left( 1\right) }=\overline{%
X}+\overline{\phi }$, $\varphi _{o}^{\left( 2\right) }=\overline{X}-%
\overline{\phi }$ for the dual ones) we get the quantum Hamiltonian (\ref
{ha3}) but, now, all the fields are chiral ones.

It is interesting to notice that the condition (\ref{blr}) is naturally
satisfied by the twisted field $\phi \left( z\right) $ of our twisted model
(TM) (see eq. (\ref{phi})).

The most convenient representation of such BS is the one in which they
appear as a product of Ising and $c=\frac{3}{2}$ BS. These last ones are
given in terms of the BS $|\alpha >$ for the charged boson and the Ising
ones $|f>$, $|\uparrow >$, $|\downarrow >$ according to (see ref. \cite{cft}
for details):
\begin{align}
|\chi _{(0)}^{c=3/2}& >=|0>\otimes |\uparrow >+|2>\otimes |\downarrow > \\
|\chi _{(1)}^{c=3/2}& >=\frac{1}{2^{1/4}}\left( |1>+|3>\right) \otimes |f> \\
|\chi _{(2)}^{c=3/2}& >=|0>\otimes |\downarrow >+|2>\otimes |\uparrow >.
\end{align}
Such a factorization naturally arises already for the TM characters \cite
{cgm4}.

The vacuum state for the TM model corresponds to the $\tilde{\chi}_{(0)}$
character which is the product of the vacuum state for the $c=\frac{3}{2}$
subtheory and that of the Ising one. From eqs. (\ref{vac1},\ref{vac3}) we
can see that the lowest energy state appears in two characters, so a linear
combination of them must be taken in order to define a unique vacuum state.
The correct BS in the untwisted sector are:
\begin{align}
|\tilde{\chi}_{((0,0),0)}& >=\frac{1}{\sqrt{2}}\left( |\tilde{\chi}%
_{(0)}^{+}>+|\tilde{\chi}_{(0)}^{-}>\right) =\sqrt{2}(|0>\otimes |\uparrow
\bar{\uparrow}>+|2>\otimes |\downarrow \bar{\uparrow}>)  \label{boud1} \\
|\tilde{\chi}_{((0,0),1)}& >=\frac{1}{\sqrt{2}}\left( |\tilde{\chi}%
_{(0)}^{+}>-|\tilde{\chi}_{(0)}^{-}>\right) =\sqrt{2}(|0>\otimes |\downarrow
\bar{\downarrow}>+|2>\otimes |\uparrow \bar{\downarrow}>) \\
|\tilde{\chi}_{((1,0),0)}& >=\frac{1}{\sqrt{2}}\left( |\tilde{\chi}%
_{(1)}^{+}>+|\tilde{\chi}_{(1)}^{-}>\right) =\sqrt{2}(|0>\otimes |\downarrow
\bar{\uparrow}>+|2>\otimes |\uparrow \bar{\uparrow}>) \\
|\tilde{\chi}_{((1,0),1)}& >=\frac{1}{\sqrt{2}}\left( |\tilde{\chi}%
_{(1)}^{+}>-|\tilde{\chi}_{(1)}^{-}>\right) =\sqrt{2}(|0>\otimes |\uparrow
\bar{\downarrow}>+|2>\otimes |\downarrow \bar{\downarrow}>) \\
|\tilde{\chi}_{(0)}(\varphi _{0})& >=\frac{1}{2^{1/4}}\left( |1>+|3>\right)
\otimes |D_{O}(\varphi _{0})>  \label{continous}
\end{align}
where we also added the states $|\tilde{\chi}_{(0)}(\varphi _{0})>$ in which
$|D_{O}(\varphi _{0})>$ is the continuous orbifold Dirichlet boundary state
defined in ref. \cite{Affleck}. For the special $\varphi _{0}=\pi /2$ value
one obtains:
\begin{equation}
|\tilde{\chi}_{(0)}>=\frac{1}{2^{1/4}}\left( |1>+|3>\right) \otimes |ff>.
\label{utgs}
\end{equation}
For the twisted sector we have:
\begin{align}
|\chi _{(0)}^{+}>& =\left( |0>+|2>\right) \otimes (|\uparrow \bar{f}%
>+|\downarrow \bar{f}>) \\
|\chi _{(1)}^{+}>& =\frac{1}{2^{1/4}}\left( |1>+|3>\right) \otimes (|f\bar{%
\uparrow}>+|f\bar{\downarrow}>).
\end{align}

Now, by using as reference state $|A>$ the vacuum state given in eq. (\ref
{boud1}), we compute the chiral partition functions $Z_{AB}$ where $|B>$ are
all the BS just obtained \cite{noi1}:
\begin{eqnarray}
Z_{<\tilde{\chi}_{((0,0),0)}||\tilde{\chi}_{((0,0),0)}>} &=&\tilde{\chi}%
_{((0,0),0)}  \label{p1} \\
Z_{<\tilde{\chi}_{((0,0),0)}||\tilde{\chi}_{((1,0),0)}>} &=&\tilde{\chi}%
_{((1,0),0)}  \label{p2} \\
Z_{<\tilde{\chi}_{((0,0),0)}||\tilde{\chi}_{((0,0),1)}>} &=&\tilde{\chi}%
_{((0,0),1)}  \label{p3} \\
Z_{<\tilde{\chi}_{((0,0),0)}||\tilde{\chi}_{((1,0),1)}>} &=&\tilde{\chi}%
_{((1,0),1)}  \label{p4} \\
Z_{<\tilde{\chi}_{((0,0),0)}||\tilde{\chi}_{(0)}>} &=&\tilde{\chi}_{(0)}
\label{p5} \\
Z_{<\tilde{\chi}_{((0,0),0)}||\chi _{(0)}^{+}>} &=&\chi _{(0)}^{+}  \label{p6} \\
Z_{<\tilde{\chi}_{((0,0),0)}||\chi _{(1)}^{+}>} &=&\chi _{(1)}^{+}.  \label{p7}
\end{eqnarray}

So we can discuss topological order referring to the characters with the
implicit relation to the different boundary states present in the system.
Furthermore these BS can be associated to different kinds of linear defects,
which are compatible with conformal invariance \cite{noi1}.

\end{document}